\documentclass[aps,prl,twocolumn,showpacs]{revtex4}
\usepackage{bbm,amsmath,amssymb,amsbsy,graphicx,times,subfigure}
\usepackage[latin1]{inputenc}
\usepackage{graphicx}% Include figure files
\usepackage{dcolumn}% Align table columns on decimal point
\usepackage{bm}% bold math

\newcommand{\virg}[1]{``#1'' }%mette fra virgolette

\begin{document}

\title{Influence of trapping potentials on the phase diagram
of bosonic atoms in optical lattices}
\date{November 4, 2004}
\author{S. M. Giampaolo, F. Illuminati, G. Mazzarella, and S. De Siena}
 \affiliation{Dipartimento di Fisica ``E. R. Caianiello'',
Universit\`{a} di Salerno, INFM UdR di Salerno, INFN Sezione di
Napoli - Gruppo Collegato di Salerno, I-84081 Baronissi (SA), Italy}

%============================= Abstract ==========================
\begin{abstract}
We study the effect of external trapping
potentials on the phase diagram of bosonic atoms
in optical lattices. We introduce a generalized Bose-Hubbard Hamiltonian
that includes the structure of the energy levels of the
trapping potential, and show that these levels are in general populated
both at finite and zero temperature.
We characterize the properties of the superfluid transition for this
situation and compare them with those of the standard Bose-Hubbard description.
We briefly discuss similar behaviors for fermionic systems.
\end{abstract}


\maketitle

{\it I) Introduction.} - 
Exciting progress in the manipulation of neutral atoms in optical
lattices has recently led, in a series of beautiful experiments,
to the first verification in atomic systems of a quantum phase
transition from a superfluid to a Mott insulator phase
\cite{ol1,ol2}. From a more general perspective, the
physics of ultracold neutral atoms in discrete structures has
become an ideal testing ground for the study and the realization
of complex systems of condensed matter physics \cite{qp1,qp2,ol4}.

Because of the great current interest in the physics of atomic
systems in optical lattices, much theoretical effort has been
devoted to the characterization of the different quantum phases
arising in these structures. Most of the studies have
been restricted to situations in which the atoms are assumed to be
confined in the lowest Bloch band of the periodic lattice potential, 
and to the ground state of the trapping harmonic oscillator
potential, so that the description is given in terms of the standard
Bose-Hubbard model \cite{fisher,qp1}. However, one observes that 
often, in practical situations, the atoms (either fermions or bosons),
while remaining confined in the lowest band of the periodic lattice potential
(for sufficiently low temperatures), nevertheless may occupy large bundles 
of excited levels of the harmonic confining potential. 
This is an indication that the description 
in terms of the standard Hubbard and Bose-Hubbard 
models is not always fully adequate, even at very low temperatures.

In the present work we aim to characterize the physics of neutral
atoms in single-band optical lattices, but including the  
energy structure due to the presence of the trapping potential. 
This leads to the introduction of generalized Hubbard and Bose-Hubbard 
models that are able to explain in a natural way the occupation 
of the excited levels of the trapping potential.
We discuss how such an energy structure 
affects the phase diagram of the system and the nature of the superfluid 
phase. The problem is rather trivial if one considers identiacl fermions,
but becomes more interesting in the case of bosons. We will comment in more
detail about fermions in the conclusions.
In the following, we consider a dilute ensemble of
identical, spinless neutral bosonic atoms subject both to a $3$-D
anisotropic harmonic trapping and a $1$-D optical lattice
potential. We derive a generalized Bose-Hubbard Hamiltonian 
and we evaluate, both numerically and analytically, the free energies 
of the system. We show that, taking into account the possibility to
populate the excited levels of the harmonic trapping potential,
allows the superfluid to become distributed among the different
energy levels. This effect of distributed superfluidity is realized
in a single, stable superfluid state, and therefore
should not be confused with the concept of multiple (fragmented), unstable 
superfluid states \cite{Nozieres}. 
This fact allows to define suitable renormalized quantities
that establish a precise mapping between the generalized and the standard
Bose-Hubbard model.

{\it II) General setting.} -
The microscopic Hamiltonian for an ensemble of bosonic atoms subject to an
optical lattice potential and confined by an additional, slowly
varying, external harmonic trapping reads
$\hat{H}=\hat{T}+\hat{V}+\hat{W}$, where $\hat{T}$ is the kinetic
energy term, $\hat{V}$ represents the external potential energy
term and $\hat{W}$ is the local two-body interaction with coupling
constant $g_{BB} = 4\pi\hbar^{2}a_{BB}/2m$, where $m$ is the mass
of the atoms and $a_{BB}$ is the $s$-wave scattering length.
In the following, we will always assume
boson-boson repulsion $a_{BB} > 0$. 
The potential energy $\hat{V}$ is made up by two different
contributions. The first contribution is an harmonic
potential $V_H$ representing
the effects of the quadrupolar trapping magnetic field,
$V_H
= m\omega^{2}
\left(x^{2}+\lambda^{2}y^{2}+\lambda^{2}z^{2}\right)/2$. Here
$\lambda$ denotes the anisotropy coefficient and $\omega$ is the
frequency associated to the harmonic trap in the $x$ direction. We
consider the situation $\lambda \ll 1$, representative of the so-called
\virg{pancake-shaped} configuration. The ground state harmonic
oscillator length in the $x$ direction is $L_x=
\sqrt{\hbar/(m\omega)}$, while for $L_y$ and $L_z$ we have that
$L_z=L_y=L_x/\sqrt{\lambda}$, i.e. $L_{y,z} \gg L_x$. The
second contribution to the potential energy is a $1$-D
periodic optical lattice $V_{opt} = V_{0}\sin^{2}\left( \pi z/a \right)$,
where $V_{0}$ is
the maximum amplitude of the light shift associated to the
intensity of the laser beam and $a$ is the lattice spacing related
to the wave vector $k$ of the standing laser light by $k=\pi/a$. 
This choice of the geometrical setting, which is not the one
commonly used in experiments, is merely due to the numerical
simplification of the energy spectra, and does not affect the general 
phenomenological features, as we will discuss in the following.

The bosonic field operators can be expanded in the basis of the
single-particle Wannier wave functions localized at each lattice
site $z_i$. The presence of the optical lattice and the strong
anisotropy define different energy scales along the different
spatial directions. The energy gap between different eigenstates 
along the $y$ direction is much smaller than the gap 
along the other ones. 
Because the typical interaction energies involved are normally not strong
enough in order to excite higher vibrational states in the $x$ and
$z$ directions, we can retain only the the lowest vibrational
state at each lattice potential well. In the harmonic
approximation, the Wannier wave functions $\Psi(\vec{r})$
factorize in the product of harmonic oscillator states $w(\vec{r})$: 
$\Psi(\vec{r}) = \sum_{i,\alpha}
\hat{a}_{i,\alpha} w(z-z_{i}) w^{\alpha}(y) w(x)$, where $z_i$ is
the center of the $i$th lattice site and $\hat{a}_{i,\alpha}$ is
the bosonic annihilation operator acting on the $\alpha$-th 
harmonic oscillator level at the $i$-th lattice site. 
Considering different geometries such as a cigar-shaped or a
fully isotropic potential will introduce extra indices in
the orthogonal ($x$, $z$) directions, thus complicating 
the numerical evaluations but not changing the basic
physical phenomena.
The local ground state spatial extension $l_z$ 
for each lattice potential well is $l_z = \left( a^4
E_R / (\pi^4 V_0 ) \right)^{1/4}$, where $E_R = (\pi \hbar)^2/2
a^{2} m$ is the lattice recoil energy.

The condition of anisotropy $\lambda \ll 1$, implies that one can
neglect in the Hamiltonian all terms proportional to powers of
$\lambda$ higher than one. Moreover, in the presence of a slowly varying
harmonic potential, we may neglect both next-to-nearest neighbor
hopping and nearest-neighbor interaction terms that are some order
of magnitude smaller than, respectively, nearest-neighbor hopping
and on-site interaction terms. The harmonic Wannier scheme thus
leads to the following multi-band Bose-Hubbard Hamiltonian
\begin{eqnarray}
\label{LatticeHamiltonian}
\hat{H} &=& \hat{H}_{l} \; \, - \;
\sum_{<i,j>,\alpha} \frac{J_\alpha}{2} \hat{a}^{+}_{i,\alpha}\hat{a}_{j,\alpha} 
\; , \\
\hat{H}_{l}\! \! &=&\! \! \sum_{i,\alpha} \left(E_0 + \lambda \alpha \hbar \omega
\right) \hat{n}_{i,\alpha} + U \! \! \! \! \!
\sum_{i,\alpha,\beta,\gamma,\delta}\! \! \! d^{\alpha,\beta}_{\gamma,\delta}
\hat{a}^{+}_{i,\alpha}\hat{a}^{+}_{i,\beta}\hat{a}_{i,\gamma}\hat{a}_{i,\delta} \; .
\nonumber
\end{eqnarray}
Here $\hat{H}_{l}$ stands for local terms. We see that neglecting
the terms proportional to $\lambda^{2}$ (and higher powers of 
$\lambda$) makes the Hamiltonian homogeneous. Thus, in this setting
(which is experimentally realistic), homogeneity is a consequence of 
strong anisotropy.
The nonlocal term in
Eq.~(\ref{LatticeHamiltonian}) is the nearest-neighbor hopping
contribution with amplitude $J_\alpha$ for a fixed harmonic energy
level $\alpha$. In standard experimental situations we
have $J_0>0$, and from the analytical expression of the hopping
amplitude one finds that $J_{\alpha+1} < J_\alpha$ for every
$\alpha> 0$. The first local term in
Eq.~(\ref{LatticeHamiltonian}) is the sum of local,
level-dependent energies proportional to the on-site number
operators $\hat{n}_{i,\alpha}$ on the $\alpha$th harmonic level on site $i$.
This term has two contributions: a zero-point energy $(E_0)$ and
an excitation energy $(\lambda \alpha \hbar \omega )$ that a boson
needs to occupy the $\alpha$th harmonic energy level. 
The second local term in the Hamiltonian is the
boson-boson on-site interaction with coupling constant $U =
g_{BB}/\left( L_x L_y l_z (2 \pi)^{3/2}\right)$. The numerical
coefficients $d^{\alpha,\beta}_{\gamma,\delta}$ read
$d^{\alpha,\beta}_{\gamma,\delta} = L_y \sqrt{2 \pi} \int_{-\infty}^{\infty}
w^\alpha(y) w^\beta(y) w^\gamma(y)w^\delta(y)dy$.
It is easy to prove that $|d^{\alpha,\beta}_{\gamma,\delta}| < 1$, with
the exception $d^{0,0}_{0,0} = 1$. Contrary to what happens in the standard
Bose-Hubbard setting \cite{ol4,qp1,qp2,sheshadri}, we see that the
local interaction term involves both inter-level interactions and inter-level 
hoppings which do not commute with the on-site number operators $\hat{n}_{i,\alpha}$
for a given harmonic oscillator level. As a consequence, the two local terms of the
generalized Bose-Hubbard Hamiltonian (\ref{LatticeHamiltonian}) do not in general 
commute. We then introduce the total on-site number operator
$\hat{N}_i=\sum_\alpha \hat{n}_{i,\alpha}$. It is
straightforward to verify that $[\hat{N}_i , \hat{H}_{l}]=0$, so
that there exists a complete set of states that are simultaneous
eigenstates of $\hat{H}_{l}$ and $\hat{N}_i$. Therefore, the
eigenstates of $\hat{H}_{l}$ can be arranged in classes characterized
by a fixed eigenvalue of $\hat{N}_i$, with each class possessing a
lowest eigenvalue. 

{\it III) Mean field theory and superfluid structure.} -
The above analysis suggests to
divide the problem of finding the
eigenstates of $\hat{H}_{l}$ in different problems in which one
determines only the eigenstates characterized by a fixed eigenvalue
$N_i$. For ease of numerical evaluation it is convenient to
adopt the grand canonical description
$\hat{K}=\hat{H}-\mu \sum_{i}\hat{N}_i$,
where $\mu$ is the chemical potential. 
%The spectrum of this Hamiltonian can be studied in
%the strong coupling regime $(U \gg J_0)$ that allows to introduce
%the effects of hopping perturbatively. 
We can analyze this problem by a mean field approach that
corresponds to the approximation in which the hopping term is
fully decoupled, in analogy with the standard Bose-Hubbard
problem \cite{sheshadri}: $\hat{a}^{+}_{i,\alpha}\hat{a}_{j,\alpha} =\left(
\hat{a}^{+}_{i,\alpha} + \hat{a}_{j,\alpha}\right)
\phi_\alpha-\phi_\alpha^2$, where $\phi_\alpha=\langle
a^{+}_{i,\alpha} \rangle=\langle a^{+}_{i,\alpha} \rangle$ is the
real (without loss of generality), homogeneous superfluid order parameter 
for the $\alpha$th harmonic energy level. The introduction of a homogeneous
superfluid parameter is justified by the homogeneity of the generalized
Bose-Hubbard Hamiltonian, which, in the mean-field approximation reads
\begin{eqnarray}
\label{MFAGrandCanonicalHamiltonian}
\hat{K}_{MF}&=&\hat{H}_{l} - \mu
\sum_{i}\hat{N}_i \nonumber \\
&-& \sum_{i,\alpha} J_\alpha \left(a^{+}_{i,\alpha}+
a_{i,\alpha}\right) \phi_\alpha + \sum_{i,\alpha}
\frac{J_\alpha}{2}
\phi_\alpha^2 \; .
\end{eqnarray}
At fixed values of both the chemical potential and the
temperature, we obtain the eigenvalues of $\hat{H}_{l} - \mu
\sum_{i,}\hat{N}_i$ associated to the eigenvectors of
$\hat{H}_{l}$. To achieve this goal we proceed by
truncating at a number $n_{max}$ of harmonic 
oscillator levels, and then diagonalizing the
$n_{max} \times n_{max}$ matrix elements of $\hat{K}_{MF}$
obtaining the eigenvalues and, hence, the free energy. We then
iterate the process by increasing $n_{max}$.
Convergence is reached when the difference between the iterated
free energies is lower than a given control 
parameter. This is a generalization of the Sheshadri approach
to the standard Bose-Hubbard problem \cite{sheshadri}.
A first physical consequence is that the mean value of the
local occupation number of the $\alpha$-th level is in general
nonvanishing, thus explaining the possible occupation of the
excited levels of the harmonic trapping potential, even at
zero temperature.
In general, the procedure outlined above
leads to consider several eigenvectors of
$\hat{H}_{l}$ and the large matrices so obtained require
thorough numerical evaluations. However, with suitable choices of
the physical parameters only the two lowest eigenstates give non
negligible contributions, and in this case the analytical
diagonalization of $\hat{K}_{MF}$ is possible. In particular, this
situation is realized when $J_0 \ll \lambda \hbar \omega$ and the
chemical potential $\mu$ is chosen in such a way that the state
$|\phi_0^{(n)}\rangle$, i.e. the eigenstate with the lowest energy
in the set of eigenstates of $\hat{H}_{l}$ with $N_i=n$, and the
state $|\phi_0^{(n+1)}\rangle$, i.e. the eigenstate with the
lowest energy in the set of eigenstates of $\hat{H}_{l}$ with
$N_i=n+1$, are nearly degenerate (i.e., the difference between the
local eigenvalues of these two states is comparable in magnitude with
$J_0$).
Let us introduce the quantity $2\Delta$, the small energy gap
between these two states. The diagonal terms of the matrix Hamiltonian
read
\begin{eqnarray}
\label{diagonalterm} 
\langle \psi_0^{(n)}|(\hat{H}_{l} - \mu
\sum_{i}\hat{N}_i ) |\psi_0^{(n)}\rangle &=& \Delta \; ,\nonumber
\\
\langle \psi_0^{(n+1)}|( \hat{H}_{l} - \mu \sum_{i}\hat{N}_i
)|\psi_0^{(n+1)} \rangle &=& - \Delta \; ,
\end{eqnarray}
while the off-diagonal terms due to the hopping are
\begin{equation}
\label{offdiagonalterm}
\! \langle \psi_0^{n} |\sum_{i,\alpha}
J_\alpha \phi_\alpha \left( \hat{a}^{+}_{i,\alpha}\! +\!
\hat{a}_{i,\alpha} \right) |\psi_0^{(n+1)}\rangle = \sum_\alpha
J_\alpha \phi_\alpha c_\alpha \; ,
\end{equation}
where $c_\alpha$ stands for $\langle \psi_0^{(n)} |\hat{a}_\alpha
|\psi_0^{(n+1)}\rangle$. The eigenvalues $\chi$ of the energy matrix
are $\chi_{1,2} = \pm \sqrt{\Delta^2+\left(\sum_\alpha
J_\alpha \phi_\alpha c_\alpha\right)^2}$. Knowing
the eigenvalues, we may write the free energy of the system
$ F=-\beta^{-1}\ln\left[2 \cosh\left( \beta \chi \right)\right]
 +\left( \sum_\alpha J_\alpha \phi_\alpha^2 \right)/2$,
with $\beta = (k_{B}T)^{-1}$, $T$ the absolute temperature, and
$k_{B}$ the Boltzmann constant. The free energy is a functional of
the whole set of order (superfluid) parameters $\phi_\alpha$
determined self-consistently through the
minimization conditions
\begin{eqnarray}
\label{minimumequation}
 \frac{\partial F}{\partial \phi_\alpha} =0 & \Rightarrow & \phi_\alpha =
 \frac{ c_\alpha}{\chi} \tanh(\beta \chi) \left(\sum_\alpha J_\alpha \phi_\alpha c_\alpha
 \right) \; .
\end{eqnarray}
Obviously, this set of self-consistent equations allows for the
existence of a disordered phase in which all the superfluid order
parameters vanish identically. The crucial question arising from
Eq.(\ref{minimumequation}) is whether ordered phases are possible
and the superfluid can be distributed among the different harmonic
levels. To have the whole superfluid concentrated on a single level,
the set of equations (\ref{minimumequation}) should allow a solution
in which only one order parameter, say $\phi_{k}$ with $k$ fixed,
is nonvanishing. Hence the question is whether a solution of this
kind is allowed. Let us first consider the case of filling factor
$n=1$, when the states considered are $|\psi_0^{(1)}\rangle$ and
$|\psi_0^{(2)}\rangle$. In the state $|\psi_0^{(1)}\rangle$, due
to the presence of exactly one atom per lattice site, the
boson-boson local interaction vanishes and hence
$|\psi_0^{(1)}\rangle$ simply factorizes in the product of
single-particle ground states $|1\rangle_{0}$, each containing one
boson. On the other hand, $|\psi_0^{(2)}\rangle$ is not an
eigenstate of $\hat{n}_{i,\alpha}$ for any $\alpha$, but it may be
written as a linear combination 
$|1,1\rangle_{\gamma,\delta}$, characterized by having
a boson in the $\gamma$th level and a boson in the $\delta$th
level: $|\psi_0^{(2)}\rangle=\sum_{\delta\le\gamma
}u_{\gamma,\delta} |1,1\rangle_{\gamma,\delta}$. The
coefficients $u_{\gamma,\delta}$ are functions of the relative
intensity of the two-body interactions and of the gap between the
different harmonic levels. Taking into
account the expression of $|\psi_0^{(2)}\rangle$ the coefficients
$c_\alpha$ that appear in the off-diagonal terms of the truncated
energy matrix Eq.(\ref{offdiagonalterm}) become equal to
$u_{0,\gamma}$ for $\gamma \neq 0$ and $\sqrt{2}\; u_{0,0}$
otherwise. From these expressions and from the form of
Eqs.~(\ref{minimumequation}), we see that the solution with
the superfluid confined in a single band is allowed if and only if
the coefficients $u_{0,\gamma}$ are zero for each $\gamma$ except
at a single fixed value $\gamma=k$. However, this
condition is inconsistent with the requirement that
$|\psi_0^{(2)}\rangle$ be an eigenstate of $\hat{H}_{l}$,
because this latter constraint implies, for every
coefficient $u_{0,\gamma'}$,
\begin{eqnarray}
\label{u0kcondition}
0&=&\left(\gamma' \lambda \hbar \omega-E_0^{(2)}\right) u_{0,\gamma'}
\, + \, 4 \sum_{\delta < \gamma} u_{\gamma,\delta}
d_{\gamma,\delta}^{0,\gamma'} \nonumber \\
&+& 2 \sqrt{2} \sum_{\gamma} u_{\gamma,\gamma} 
d_{\gamma,\gamma}^{0,\gamma'} \; ,
\end{eqnarray}
where $E_0^{(2)}$ is the eigenvalue of $\hat{H}_{l}$ associated
to $|\psi_0^{(2)}\rangle$. For any $\gamma'$ such that $k+\gamma'$
is even (this assumption is important because if $k+\gamma'$ is
odd, then $d_{0,\gamma'}^{0,k}=0$), solving for $u_{0,k}$ yields 
\begin{equation}
\label{u0kcondition2}
u_{0,k}=\frac{-1}{4  d_{0,\gamma'}^{0,k}}\! \!
\left( 4 \! \! \!  \sum_{0<\delta< \gamma}\! \! \! 
u_{\gamma,\delta} d_{\gamma,\delta}^{0,\gamma'}\!+2 \sqrt{2} \sum_{0<\gamma} u_{\gamma,\gamma}
d_{\gamma,\gamma}^{0,\gamma'}\! \right) \; .
\end{equation}
We thus arrive at a set of equations with a number of
variables less than the number of equations. Because the extra
equations are linearly independent from each other it is
impossible to find a set of hamiltonian parameters that allows
$u_{0,\gamma}=0$ for any $\gamma \neq k$.
Numerical evidence shows that qualitatively similar results hold for $n>1$.
In the case $n=0$ (hard core limit),
the superfluid is confined to the harmonic oscillator ground state.
This result may be explained by observing that in
this case the local interactions do not play any role.

\begin{figure}
\centering
\includegraphics[width=8cm]{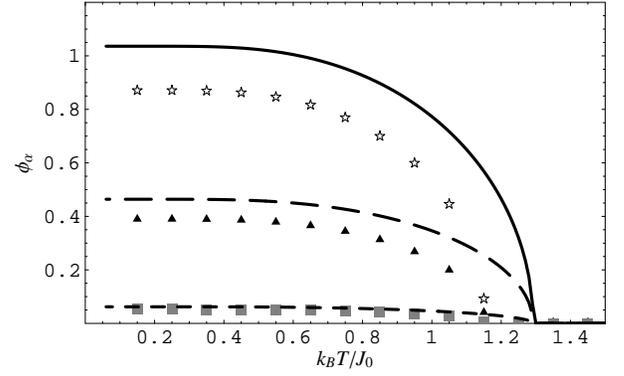}
\caption{From top to bottom: the first three nonvanishing 
superfluid order parameters $\phi_{0}$, $\phi_{2}$, and $\phi_{4}$ 
as a function of the temperature $k_{B} T/J_0$, for
$\lambda \hbar \omega=0.1U$ and $J_{\alpha} - J_{\alpha +1}=0.001 J_0$. 
Lines for $\Delta=0$, points for $\Delta=0.7 J_0$. Notice the existence
of a unique critical temperature at fixed values of $\Delta$.} 
\label{parameterfig}
\end{figure}

{\it IV) Results and discussion.} -
In Fig.~\ref{parameterfig} we show the behavior of the superfluid order 
parameters associated to different harmonic oscillator energy levels
as functions of the temperature 
for different values of the energy gap between the states
$|\psi_0^{(2)}\rangle$ and $|\psi_0^{(1)}\rangle$ (In all figures
the quantities plotted are dimensionless).
From Fig.~\ref{parameterfig} we see that the ratio of
superfluid in the different levels is fixed and independent of the
temperature. This is clear evidence that, although the superfluid
is distributed among the different levels, the superfluid state is
unique, and therefore stable \cite{Nozieres}.
From Eq.~(\ref{minimumequation}) the ratios are simply: $r_\alpha= c_\alpha /
\sqrt{\sum_\alpha c_\alpha^2}$, where $r_\alpha$ is the fraction of
superfluid in the level $\alpha$. The fact that the superfluid state
is unique leads naturally to
the existence of a unique critical temperature $T_c$. The
latter depends on the energy gap $\Delta$ and on the
different hopping amplitudes $J_\alpha$, and 
may be obtained from the set of Eqs.~(\ref{minimumequation}): 
it is the temperature for which there exists a double
solution at $\phi_\alpha=0$ for every $\alpha$:
\begin{equation}
\label{criticaltemperature}
\beta_c \Delta = \arctan \! {\mbox h} 
\left(\frac{\Delta }{\sum_\alpha c_\alpha^2 J_\alpha}\right) \; .
\end{equation}
\begin{figure}
\centering
\includegraphics[width=8cm]{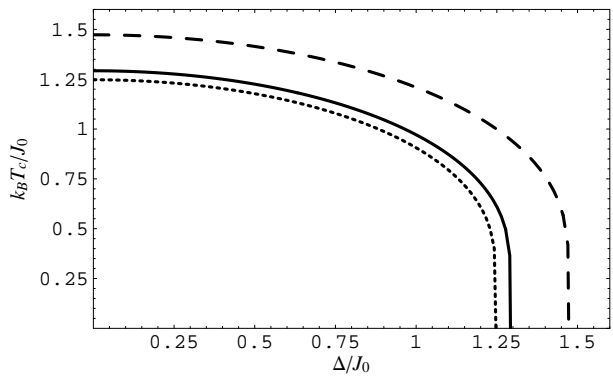}
\caption{Critical temperature $k_B T_c/J_0$ as function of the gap
$\Delta/J_0$ for different values of $\lambda \hbar \omega$ at
fixed $J_\alpha-J_{\alpha+1}=0.001J_0$. Dashed, solid, and
dotted lines are respectively for 
$\lambda \hbar \omega = 0.2U, \, 0.1U, \, 0.01U$.}
\label{tc1fig}
\end{figure}
%\begin{figure}
%\centering
%\includegraphics[width=8cm]{tc2bw.eps}
%\caption{Critical temperature $k_B T_c/J_0$ as function of the gap
%$\Delta/J_0$ for different values of $J_\alpha-J_{\alpha+1}$ at
%fixed $\lambda \hbar \omega=0.1U$.
%Solid, dashed, and dotted lines are respectively for 
%$J_\alpha-J_{\alpha+1} = 0.2 J_0, \, 0.1 J_0, \, 0.001 J_0$.}
%\label{tc2fig}
%\end{figure}
\begin{figure}
\centering
\includegraphics[width=8cm]{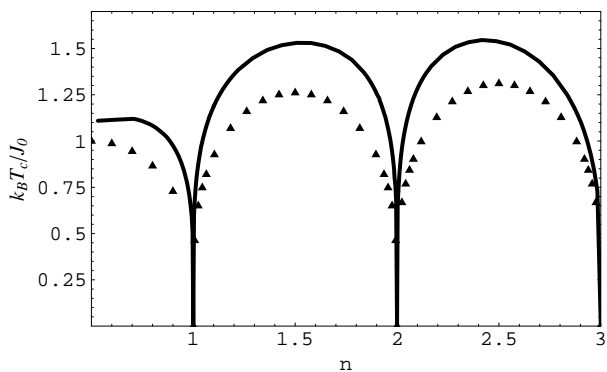}
\caption{Critical temperature $k_B T_c/J_0$ as a 
function of the filling factor $n$
for $J_\alpha-J_{\alpha+1}=0.001J_0$ and
$\lambda \hbar \omega=0.1U$. The dots correspond to 
the analytical solution obtained with the two lowest
local states. The full line corresponds to the numerical
solution including the first $40$ local states.} 
\label{tcrhofig}
\end{figure}

In Fig.~\ref{tc1fig} we show the behavior of
$T_c$, for varying $\lambda \hbar \omega$ at fixed
$J_\alpha$. As far as $T_{c}$ is concerned,
one can see from this figure and Eq.~(\ref{criticaltemperature}) 
that the transition is analagous to that of a 
standard Bose-Hubbard model with renormalized hopping amplitude
$J_R=\sum_\alpha c_\alpha^2 J_\alpha$. In Fig.~\ref{tcrhofig} we
show the behavior of the critical temperature as a function of the
filling factor $n$ both for the analytical solution with only the
two lowest-lying local states and the numerical solution obtained
after full convergence is reached.
The zero-temperature quantum phase transition from a superfluid to
a Mott-insulator phase is recovered for the
value $\Delta_c$ at which $T_c = 0$. From
Eq.~(\ref{criticaltemperature}) one has $\Delta_c = \sum_\alpha
c_\alpha^2 J_\alpha$.

In conclusion, we have studied the properties of an ensemble of
neutral bosonic atoms in an optical lattice, including 
the energy structure due to the presence of a superimposed
trapping potential.
We have shown that the system possesses an ordered phase 
in which the superfluid is distributed among the different 
harmonic energy levels. Our analysis holds in a similar
way in the much simpler instance of spin-polarized fermions
and thus provides the physical mechanism explaining 
the population of the excited levels of the
harmonic trapping potential. Work is in progress
for the case of interacting fermions, and will be reported
elsewhere. An interesting possibility for future research lies
also in the study of the energy structure for multicomponent
systems and mixtures of bosonic and fermionic atoms in optical 
lattices.
We thank INFM, INFN, and MIUR under national project PRIN-COFIN
2002 for financial support.

\end{document}